\documentclass{WileyMSP-template}
\usepackage{graphicx}
\usepackage{epstopdf}
\usepackage{amsmath}
\usepackage{color}

\newcommand{\newpara}
    {
    \vskip 0.3cm
    }
\newcommand{\SL}[1]{\textcolor{black}{#1}}
\graphicspath{{C:/Users/Drshi/Documents/article_UOttawa_Poly/Article_in_FemtoQ/Fractional_SPM/Submission/LPR/LaTeX-template/MSP-Latex-Template/Figure/}}
\begin{document}

\pagestyle{fancy}
\rhead{\includegraphics[width=2.5cm]{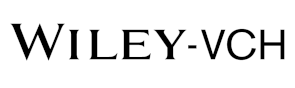}}

\title{Experimental emulator of pulse dynamics in fractional \\ nonlinear Schr\"{o}dinger equation}
\maketitle

\author{Shilong Liu$^+$,}
\author{Yingwen Zhang,}
\author{St\'{e}phane Virally,}
\author{Ebrahim Karimi,}
\author{Boris A. Malomed,}
\author{and Denis V. Seletskiy $^*$}


\begin{affiliations}
Shilong Liu\\
femtoQ Lab, Department of Engineering Physics, Polytechnique Montr\'{e}al, Montr\'{e}al,
Qu\'{e}bec H3T 1J4, Canada\\
Tempo Optics Inc., Montr\'{e}al, Qu\'{e}bec H3T 1W9, Canada\\
$^+$ dr.shilongliu@gmail.com\\
\newpara
Yingwen Zhang, Ebrahim Karimi\\
Department of Physics, University of Ottawa, 25 Templeton, Ottawa, Ontario, K1N 6N5 Canada\\
National Research Council of Canada, 100 Sussex Drive, Ottawa, K1A0R6, Canada\\
\newpara
St\'{e}phane Virally\\
femtoQ Lab, Engineering Physics Department, Polytechnique Montr\'{e}al, Montr\'{e}al,
Qu\'{e}bec H3T 1J4, Canada\\
\newpara
Boris A. Malomed\\
Department of Physical Electronics, Faculty of
Engineering, and Center for Light-Matter Interaction, Tel Aviv University, Tel Aviv 69978, Israel
Instituto de Alta Investigaci\'{o}n, Universidad de Tarapac\'{a},
Casilla 7D, Arica, Chile\\
\newpara
 Denis V. Seletskiy\\
femtoQ Lab, Engineering Physics Department, Polytechnique Montr\'{e}al, Montr\'{e}al,
Qu\'{e}bec H3T 1J4, Canada\\
$^*$ denis.seletskiy@polymtl.ca\\
\end{affiliations}





\begin{abstract}
We present a nonlinear optical platform to emulate a nonlinear \textit{L\'{e}vy waveguide} that supports the pulse propagation governed by a generalized
fractional nonlinear Schr\"{o}dinger equation (FNLSE). Our approach distinguishes between intra-cavity and extra-cavity regimes, exploring
the interplay between the effective fractional group-velocity dispersion (FGVD) and Kerr nonlinearity. In the intra-cavity configuration, we
observe stable \textit{fractional solitons} enabled by an engineered combination of the fractional and regular dispersions in the fiber cavity.
The soliton pulses exhibit their specific characteristics, \textit{viz.}, ``heavy tails" and a ``spectral valley" in the temporal and frequency
domain, respectively, highlighting the effective nonlocality introduced by FGVD.
Further investigation in the extra-cavity regime reveals the
generation of spectral valleys with multiple lobes, offering potential applications to the design of high-dimensional data encoding.  \SL{To elucidate the spectral valleys arising from the interplay of FGVD and nonlinearity, we have developed an innovative ``force" model supported by comprehensive numerical analysis. }These findings open new avenues for experimental studies of spectral-temporal dynamics in fractional nonlinear systems.
\end{abstract}

\section{Introduction}

The fractional nonlinear Schr\"{o}dinger equation (FNLSE) has drawn much interest to possibilities of its
implementation in fractional quantum mechanics and nonlinear fractional physics
\cite{laskin2000fractionalpath,longhi2015fractional,zhang2015propagation,Boris2024Basic}. Within this
framework, a particularly interesting topic is the study of the pulse (quasi-soliton) dynamics governed by
FNLSE, where the fundamentally novel feature is to address the fractional derivative in the governing
equations \cite{atangana2016new,kee2024memory,chen2024three}.
The effective nonlocality introduced by the fractional derivative results in novel
modal shapes, such as the fractional solitons \cite{fujioka2010fractional},
~which essentially expands the traditional concept of self-trapping and solitary waves. Different from the
traditional solitons that are secured by the balance between the usual (non-fractional) dispersion and
nonlinearity \cite{agrawal2000nonlinear}, fractional solitons are stabilized by the interplay of the
nonlinearity with the fractional nature of dispersion. The fractional solitons usually feature pronounced
``heavy tails", thus enhancing the interaction range beyond the immediate vicinity of the waveforms
\cite{fujioka2010fractional,chen2018optical,qiu2020soliton}.
\SL{Many numerical works have revealed diverse modes predicted by FNLSE, including fractional gap solitons
\cite{Dong1}, vortex \cite{Pengfei2}, quadratic \cite{quadratic}, and spin-orbit coupling
\cite{sakaguchi2022one}. In addition, non-trivial beam propagation dynamics were studied within FNLSE, such as breathers \cite{zhang2015propagation,zhang2019anomalous,tan2024manipulating}, symmetry-breaking \cite{zhang2016pts,chen2020spontaneous,ZhongSSB2023,he2024symmetry}, bifurcations \cite{li2021symmetry,strunin2023symmetry}, and frequency conversion \cite{li2023second}. }

\newpara
In spite of the extensive theoretical developments, experimental implementation of pulse dynamics in
fractional nonlinear systems remains a challenge. The main difficulty is to find a nonlinear \textit{L\'{e}vy
waveguide}~(LW), which could implement the effect of both fractional group-velocity dispersion (FGVD)
and nonlinearity. \SL{Recently, we have reported the realization of a linear optical LW in the spectral-temporal domain, where the FGVD is
determined by an adjustable \textit{L\'{e}vy index} (LI), $\alpha \in [0, 2]$ \cite{liu2023experimental}. By applying a fractional dispersion phase through a pulse shaper system, unique pulse dynamics, such as pulse splitting and merging, were observed within the framework of the fractional Schr\"{o}dinger equation (FSE).}
This result suggests the possibility of using the ubiquitous Kerr nonlinearity of optical materials and FGVD to emulate nonlinear LWs.
\newpara
In this work, we engage a pulse shaper to the fiber laser cavity, building a nonlinear LW to emulate the pulse
 dynamics governed by FNLSE. We establish two distinct operational regimes, intra- and
extra-cavity ones. In the intra-cavity regime, we observe fractional solitons due to the main interplay of the
FGVD and Kerr nonlinearity. We thus identify a stability boundary, determined by the  FGVD
length and  LI. Typically, a valley is observed in the spectrum of the fractional solitons. To further
investigate and enhance the spectral valley, we implement an extra-cavity regime, in which we can adjust
both the FGVD and nonlinear strength. Furthermore, we have designed a segmented fractional phase
profile in the initial pulse, which makes it possible to produce a spectral valley with multiple spectral lobes
in the extra-cavity regime. The pulse shaper used in the current setup is also known as efficient tools for performing spectral-temporal
shaping in the linear pulse shaper settings
\cite{xu2004controlling,monmayrant2010newcomer,weiner2011ultrafast}, supercontinuum tailoring
\cite{dudley2006supercontinuum,hary2023tailored}, as well as photonic-AI
\cite{iegorov2016direct,ma2021deep}.
\newpara
The spectral valley is a typical feature resulting from the interplay between the FGVD and nonlinearity.
Similar behavior was observed in the presence of regular (non-fractional) GVD and nonlinearity, in the
form of self-phase modulation (SPM).
\SL{These structures are crucially important for the formation of soliton pairs \cite{akhmediev1997multisoliton,herink2017real,Liu2022SM}, high-order soliton \cite{kivshar2003optical,agrawal2000nonlinear,karpman1993radiation}, spectral-temporal pulse shaping
\cite{mollenauer1980experimental,stolen1978self,boscolo2017shaping,tzang2018adaptive,ansari2018tailoring,zhang2024dissipative},
spatial-temporal beam coupling \cite{wright2022physics,shen2023roadmap}, supercontinuum generation \cite{dudley2006supercontinuum} and few-cycle laser \cite{apolonski2000controlling,riek2017subcycle}.}
They are also observed in self-similar and fractal patterns
\cite{fermann2000self,dudley2007self,biesenthal2022fractal,wu2023synchronization}. In these contexts,
the progression in the number $N$ of the generated spectral lobes follows a \textit{cascading} sequence
with increasing nonlinearity, such as $1 \rightarrow 2 \rightarrow \cdot \cdot \cdot \rightarrow N$,
determined by the accumulated nonlinear phase shift $\approx (N - 0.5) \pi$
\cite{stolen1978self,agrawal2000nonlinear}. In contrast, the spectral valleys reported in the present work
are connected by a sequence of \textit{direct} transitions at the same level of nonlinearity, i.e., $1
\rightarrow N$, which is an advantage of the reported scenario. These spectral-valley modes are
demonstrated to be efficient for implementing high-dimensional data encoding.


\section{Pulse dynamics in FNLSE (Fractional Nonlinear Schr\"{o}dinger Equation)}
The generalized FNLSE for a slowly-varying amplitude $\Psi(z,t)$ of the optical pulse propagating in the nonlinear LW is \cite{liu2023experimental,malomed2024fractional}
\begin{equation}\label{FNLSE}
  i\frac{\partial \Psi}{\partial z}=\left[\frac{ D}{2} \left( \frac{-\partial^2}{\partial
  t^2}\right)^{\frac{\alpha}{2}}+\frac{ \beta_2}{2}  \frac{\partial^2}{\partial t^2}+i\frac{
  g}{2}\right]\Psi- \gamma|\Psi|^2\Psi,
\end{equation}
where $z$ and $t$ are the propagation distance and  reduced time
\cite{agrawal2000nonlinear,dudley2006supercontinuum}, respectively. On the right-hand side, the first
term is the fractional (Riesz) time derivative \cite{bayin2016definition,liu2023experimental}, with LI
$\alpha$ and coefficient $D$. The second term represents the usual second-order GVD with coefficient $
\beta_2$. $g$ represents the gain coefficient, and $\gamma$ defines the strength of the SPM
nonlinearity.
\newpara
The Riesz derivative is defined, with the help of the Fourier transform, as the nonlocal operator acting in
the temporal domain:
\begin{equation}\label{Nonlocal}
 \left( \frac{-\partial^2}{\partial t^2}\right)^{\frac{\alpha}{2}}
 \Psi(t)=\frac{1}{2\pi}\int_{-\infty}^{+\infty} |\omega|^{\alpha}d\omega
 \int_{-\infty}^{+\infty}dt'e^{-i\omega(t-t')}\Psi(t').
\end{equation}
\begin{figure}[!ht]
  \centering
  \includegraphics[width=19cm]{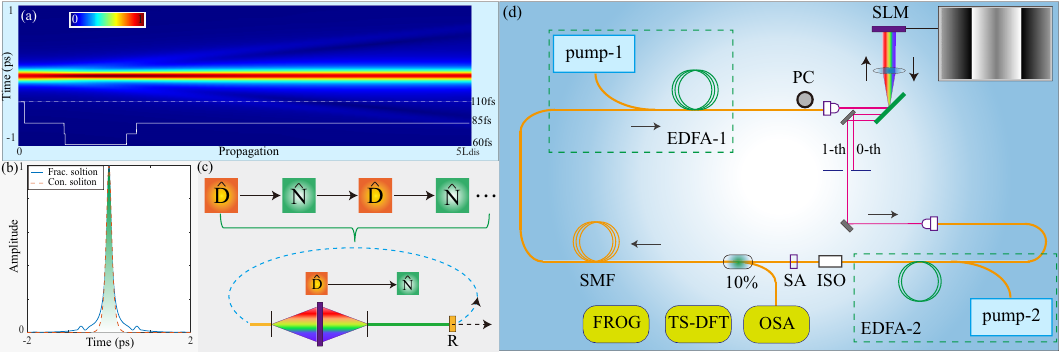}
  \caption{Simulations for the pulse dynamics in the framework of FNLSE, and
its implementation in optics. (a) The pulse-intensity evolution in five
FDLs, where the initial pulse is set to be an L\'{e}vy distribution, $A\cdot
\mathcal{L}_{1.2,1}(t/T_{0})/|\mathcal{L}_{1.2,1}(t/T_{0})|$, see Eq. (%
\protect\ref{E-Levy}). Here, $T_{0}$ is $60.1$ fs, $A=18.76$, and $\protect%
\alpha =1.2$. The inserted fragmented white line shows the half-pulse width,
which is $85$ fs in the output. (b) The pulse amplitude for the produced
fractional soliton with $\protect\alpha =1.2$ and the conventional soliton
(the dashes red profile) are displayed for comparison. (c) The scheme for the
implementation of FNLSE in optics: the extra-cavity setup ($R=0$), and the intra-cavity one ($R>0$). (d)
The experimental setup of the intra-cavity is designed to emulate FNLSE. Here the
labels are defined as follows: pump-1(2): the pump laser operating at $980$
nm; \SL{EDFA-1(2): erbium-doped fiber amplifiers}; PC: the polarization controller; SLM: the
spatial light modulator; ISO: the fiber isolator; SA: the saturable absorber; SMF: the single-mode fiber.}\label{F1}
\end{figure}

The split-step algorithm can be used to solve the generalized FNLSE (Eq. \ref{FNLSE}). Regarding the
dispersion part, the equation for marching from the segment $n$ to $n+1$ is
\begin{equation}\label{Freq}
  \tilde \Psi_{n+1}(\omega,z+\Delta z)= \tilde\Psi_{n}(\omega,z)\, \exp\left[-i \frac{ D}{2}
  |\omega|^\alpha \Delta z+i{\frac{ \beta_2}{2}\omega^2}\Delta z \right].
\end{equation}
The nonlinearity effect of SPM is implemented as
 \begin{equation}\label{SPM}
   \Psi_{n+1}(t,z+\Delta z)= \Psi_{n}(t,z)\,\exp{(iBI(t))},
 \end{equation}
where $B=\gamma P_k \Delta z$ represents the phase contribution from SPM. $I(t)$ represents the
normalized temporal intensity (TI) of the pulse; $\gamma$ and $P_k$ indicate the Kerr coefficient and the
peak power, respectively.
Based on this model, the pulse propagation in the nonlinear LW is determined by the regular and fractional
GVD and SPM. Three cases are considered in the setting
which makes it possible to neglect the linear gain: (i) the regular NLSE which does not include the
fractional GVD ($D=0$); (ii) the pure FNLSE which does not include regular GVD ($\beta_2=0$); and (iii)
the generalized FNLSE which includes both the regular and fractional GVDs.
\newpara
In case (i), NLSE has commonly known solutions for fundamental and higher-order solitons
\cite{agrawal2000nonlinear}. In case (ii), pure FNLSE supports stable pure ``fractional soliton" solutions
for $\alpha \in (1, 2)$ \cite{Boris2024Basic}. In case (iii), the generalized FNLSE  can support more
complex solutions combining fractional and regular dispersion.

To obtain the solution for the fractional soliton, we consider the initial
pulse with a \textit{L\'{e}vy distribution} as $\Psi(t, z=0)=\mathcal{L}_{\alpha' ,\gamma }(t)$:

\begin{equation}
\mathcal{L}_{\alpha' ,\gamma }(t)=\frac{1}{\pi }\int_{0}^{\infty }e^{-\gamma
q^{\alpha' }}\cos (tq)\,dq,  \label{E-Levy}
\end{equation}%

where $\gamma > 0$ and $\alpha' \in (0, 2]$. For $\alpha' = 1$, the distribution is equivalent to the
classical Cauchy distribution, and for $\alpha' = 2$, it is tantamount to the Gaussian distribution but with
fatter tails \cite{ismail2013novel}.
\newpara
Figure \ref{F1}(a) displays the pulse dynamics for the pure FNLSE with $\alpha = 1.2$ \SL{by setting $\alpha'=\alpha$}, which does
not include regular GVD, over the propagation distance equivalent to five fractional dispersion lengths
(FDL), with FDL defined as
\begin{equation}
  L_{\alpha, \mathrm{FDL}} = \frac{2 T_0^\alpha}{|D| \alpha}, ~~~~ \alpha \in (0,2],
\end{equation}
where $T_0$ is the duration of the initial pulse. \textbf{More details about the FDL are shown in
Supplement 1}. The initial pulse with the L\'{e}vy distribution defined in \eqref{E-Levy}, eventually forms a stable solution, as evidenced by monitoring its energy and pulse width. The blue solid line in Fig.
\ref{F1}(b) represents the final fractional soliton pulse, which exhibits the feature of heavy tails.
For reference, the dashed line shows the conventional soliton shaper, $\text{sech}(t)$, as produced by the regular NLSE.

\newpara
Realizing the pure FNLSE poses challenges in the experiment, while achieving a generalized FNLSE, as
defined by  \eqref{FNLSE}, is more feasible. The cavity periodically applies transformations  $\hat{D}$
and $\hat{N}$, induced by the GVD and nonlinearity, respectively, to the laser pulse circulating in the
cavity, as shown in  Fig. \ref{F1}(c), thereby creating an effective fractional nonlinear system which governs
the pulse evolution. This process is similar to generating ``split-step solitons" \cite{driben2000split} and
``pure-quartic solitons" \cite{runge2020pure}. In the regime with weak dispersion, the pulse evolution may
be approximated by a single-pass structure \cite{agrawal2000nonlinear}. Thus, this setup operates in
the ``extra-cavity" regime ($R = 0$), as opposed to the  ``intra-cavity" one ($R > 0$)\SL{, as shown in Fig. \ref{F1}(c)}.
\newpara
In our demonstration, we emulate the nonlinear LW using the mode-locked fiber laser (MLFL) and pulse
shaper. In the ``intra-cavity" configuration, the pulse shaper is placed inside the cavity to provide and
maintain both regular and fractional dispersion, while the optical fiber and amplifier induce nonlinearity.
In the ``extra-cavity" configuration, the pulse shaper is moved outside the cavity, to act on the soliton
pulse from MLFL. This configuration allows independent control of the action of the FGVD and nonlinearity on the
incident pulse. In the experiments, we observed a stable fractional soliton pulse in the MLFL,
demonstrating the efficiency of this approach to emulating pulse dynamics described by the generalized
FNLSE.

\section{Results}

\subsection{Intra-cavity regime: fractional soliton}

\begin{figure}
    \centering
    \includegraphics[width=19cm]{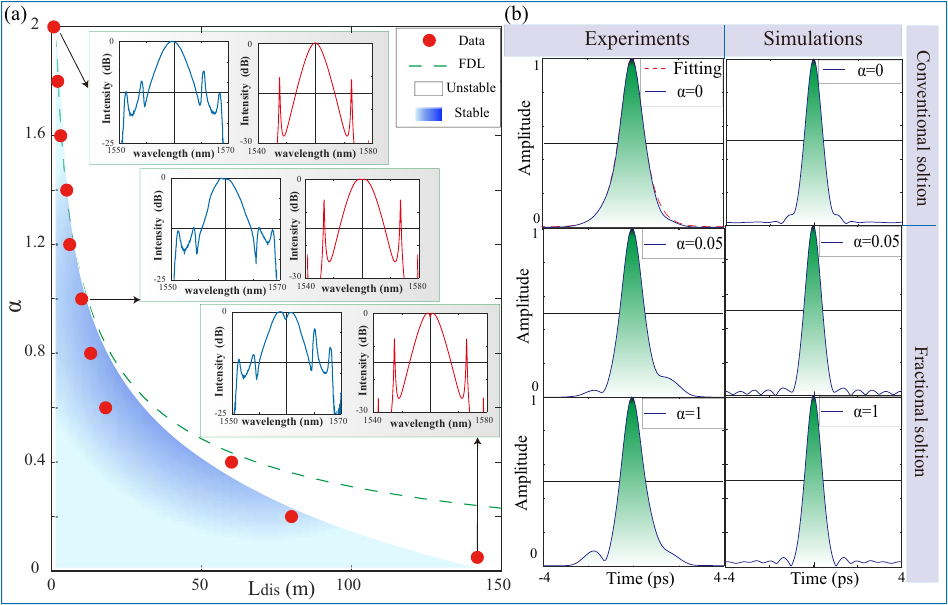}
   \caption{The spectral-temporal analysis of the soliton pulses produced by
the mode-locked fiber laser. (a) The stability analysis produced by varying
LI ($\protect\alpha $) and dispersion length ($L_{\mathrm{dis}}$). Data
points indicate the boundary between stable and unstable soliton states; the
dashed line represents the fractional-dispersion length (FDL) for $T_{0}=250$ fs.
Insets show the measured and simulated (left and right sides, respectively)
spectra for specific parameter settings: [$\protect\alpha $, $L_{\mathrm{dis}%
}$] = [$0,0$], [$1,10$ m], and [$0.05,142$ m], respectively. (b) The comparison of the
pulse amplitude produced by the experiment and simulations. The experimental profiles
were reconstructed by the FROG system.}
    \label{F2}
\end{figure}

Figure \ref{F1}(d) shows the intra-cavity setup, which divides the cavity into free-space and in-fiber
sections. The free-space one incorporates a $4f$ pulse shaper, equipped with a Spatial Light Modulator
(SLM) that is loaded with a regular and fractional phase, as defined in \eqref{Freq}.  SLM is responsible for
inducing the fractionality in the cavity, by imparting the corresponding phase shift to the
passing pulse utilizing a properly designed hologram (schematically shown
on the right-hand side of SLM in Fig. \ref{F1}(d)), as previously realized
in the linear system \cite{liu2023experimental}. The fiber component consists of two identical Er-doped
fiber amplifiers (EDFA) and a single-mode fiber (SMF), which enhance the pulse energy and furnish the
nonlinearity and regular dispersion. A saturable absorber (SA) is inserted into the fiber cavity to provide
the saturation necessary for stable mode-locking.  The output coupler diverts $\approx$ 10\% of the pulse
energy for monitoring purposes. The net dispersion in the cavity is $\sim$ -0.12 $\mathrm{ps^2}$, which ensures that the system readily produces the conventional soliton \cite{grelu2012dissipative,Liu2022SM}.
For diagnostics, we employ Frequency-Resolved Optical Gating (FROG) and an Optical Spectrum Analyzer
(OSA) to measure the temporal and spectral profiles, respectively. Additionally, a Time-Stretch Dispersive
Fourier Transform (TS-DFT) system is used to assess the pulse-by-pulse stability of the generated pulse
train.
\newpara
Figure \ref{F2} (a) presents the spectral-temporal analysis of the fractional solitons produced in the
intra-cavity setup, highlighting the effect of FGVD, as defined by two key parameters:  LI ($\alpha$) and
dispersion distance $L_{\mathrm{dis}}$. To explore their effects on the output soliton, $\alpha$ was
varied from 0 to 2, and $L_{\mathrm{dis}}$ ranges from 0 to 150 m, with a fixed dispersion coefficient
$D$ of $-21 \times 10^{-3} \text{ ps}^\alpha/\text{m}$. Typically, when the flat phase ([$\alpha$,
$L_{\mathrm{dis}}$]=[0, 0]) is loaded into the pulse shaper (PS), the cavity produces stable conventional
solitons. The spectrum of this configuration is shown in the upper left panel, and the corresponding
spectrum produced by the simulations is shown in the right panel. The simulation method for the setup is
similar to one employed in our studies of the  ``soliton molecules" (bound states of solitons) in the fiber
laser cavity \cite{Liu2022SM} (\textbf{also see more details below in Method-1}).

\newpara
The experimental and simulated spectra for the fractional-phase settings
with [$\alpha=1$, $L_{\mathrm{dis}}$=10 m] and [$\alpha=0.05$, $L_{%
\mathrm{dis}}$=142 m] are shown in the middle and lower parts of
Fig. \ref{F2}(a), respectively. Compared to the traditional solitons, the
fractional ones exhibit significant spectral alterations (most notably, a
dip at the spectrum's center), modifying the pulse profile. To
examine their temporal intensity, we employed the FROG system to reconstruct
the spectral-temporal profiles. Figure \ref{F2}(b) presents the so
reconstructed temporal profiles. The top left panel exhibits the intensity
of the conventional soliton, with the $\text{sech}(T)$ shape, as indicated
by the fitting dashed line. In contrast, two subsequent panels show the
fractional pulses for the settings with [$\alpha=1$, $L_{\mathrm{dis}}$=10 m]  and [$\alpha=0.05$,
$L_{\mathrm{dis}}$=142 m], respectively.
The characteristic ``heavy tails" of the pulses are clearly
observed.
\newpara
The observed modifications are supported by simulations displayed in the
right panels. This characteristic behavior is also observed for $\alpha \in
(1,2)$; additional results reconstructed \ by the FROG measurements can be
found in \textbf{Supplement 2}. These features are emblematic of fractional
solitons, being consistent with the characteristics reported in previous
theoretical studies \cite%
{fujioka2010fractional,yao2018solitons,malomed2021optical}.

Based on Fig. \ref{F2}(a), for given LI $\alpha $, it is observed that the
output remains stable as $L_{\mathrm{dis}}$ increases, up to a certain
threshold value beyond which the cavity becomes unsuitable. The reason is that larger
$L_{\mathrm{dis}}$ induces the larger FGVD, leading to exceeding the temporal window of the PS and
thus producing a distortion in the pulse profile \cite{weiner2011ultrafast}.
In Fig. \ref{F2}(a), the left blue area represents stability regions for the soliton
pulse, while the right blank area indicates instability, where a robust
spectrum does not form. The stability boundary aligns with the variation of
the FGVD length as defined above and is shown here by the green dashed line.
\newpara
A noteworthy feature revealed by the experiments is that the stability
range of the output fractional soliton pulses extends, in terms of LI $%
\alpha $, from $0$ to $2$. In contrast, the predicted stability area for the
pure FNLSE is found for $\alpha $ between $1$ and $2$, as it gives rise to
the collapse at $\alpha \leq 1$ \cite{Boris2024Basic}. The reason for this
difference is that the constructed cavity is not a pure FNLSE system, as it
also incorporates regular GVD and other effects, such as saturable
absorption. These effects help to suppress the destabilizing factor of the
collapse and extend the stability area to $\alpha \leq 1$.  As a result, the intra-cavity system is closer to the
generalized FNLSE model.

\subsection{The extra-cavity FNLSE-emulating regime: spectral valleys}
Limited by the stability boundary in the intra-cavity regime, as illustrated
in Fig. \ref{F2}(a), the cavity is not stable for the larger FGVD, and thus
the depth of the spectral dip cannot be made more pronounced. To circumvent
this limitation, we ran the experiments with the extra-cavity regime. For
this purpose, we relocated the PS outside the cavity and
incorporated EDFA to control the nonlinearity strength, by varying the
EDFA's pump power is from $70$ to $350$ mW. In this regime, the pulse shaper is
used to change the action of both the regular and fractional GVD on the
initial pulse. The extra-cavity setup offers the flexibility to alter the
initial conditions and nonlinearity. Technical details concerning the
extra-cavity setup are available in \textbf{Supplementary 3}.
\begin{figure}[!h]
    \centering
\includegraphics[width=16cm]{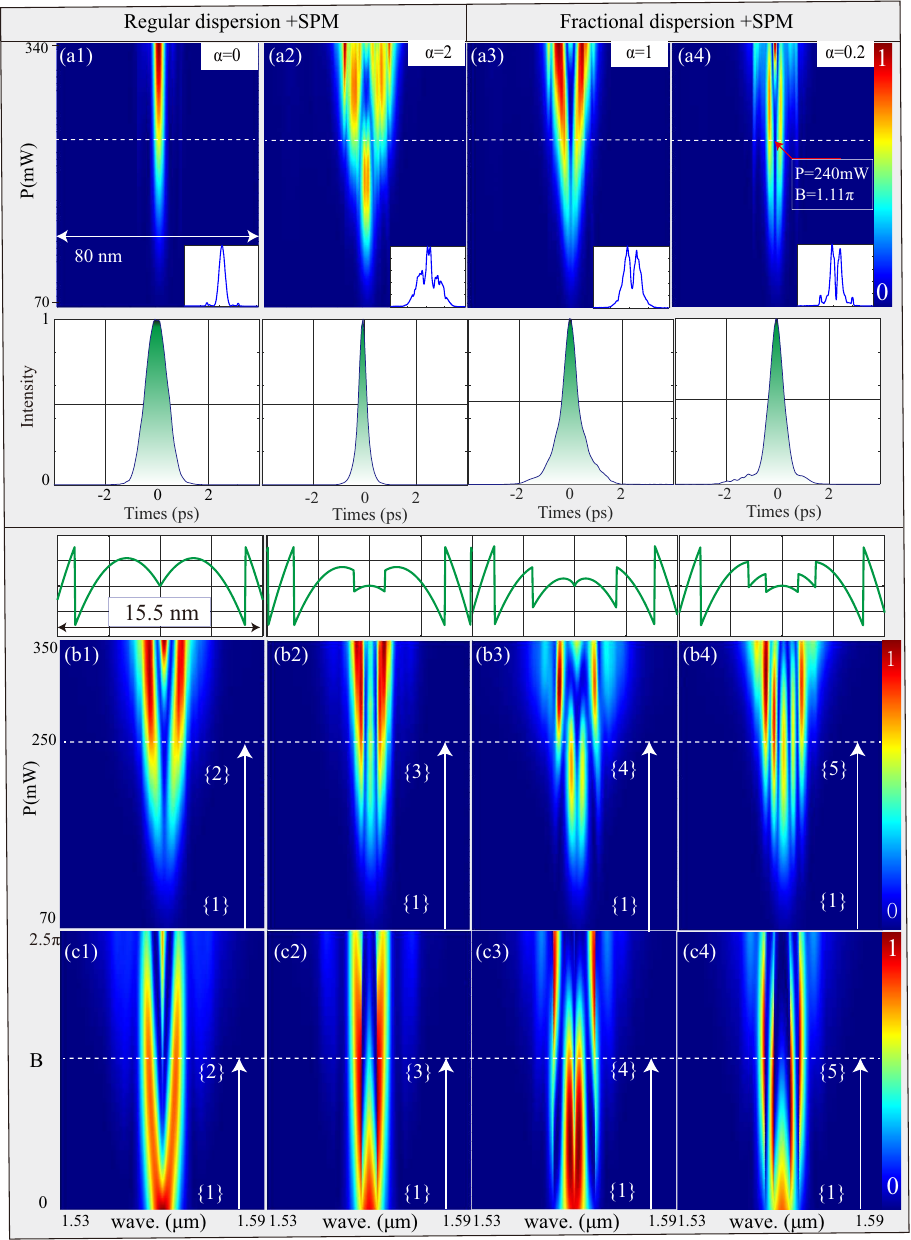}
    \caption{Spectral responses to varying the EDFA's pump power in the
extra-cavity regime. (a1)-(a4): Recorded spectra produced by the flat [$%
\protect\alpha =0,L_{\mathrm{dis}}=0$], second-order [$\protect\alpha =2,L_{%
\mathrm{dis}}=1.7L_{\protect\alpha ,\mathrm{FDL}}$], fractional-order [$\protect\alpha %
=1,L_{\mathrm{dis}}=2.42L_{\protect\alpha ,\mathrm{FDL}}$] and [$\protect%
\alpha =0.2,L_{\mathrm{dis}}=0.88L_{\protect\alpha ,\mathrm{FDL}}$] GVD,
respectively. The bottom panels show the temporal intensity reconstructed by
the FROG system for the pump power in 240 mW. (b1)-(b4): Showcasing the spectral valleys with multiple
lobes, under the action of the segmented fractional phase, as indicated at
the top of each panel and defined in Eqs. \protect\ref{E7}- \protect\ref{E10}%
, respectively. (c1)-(c4): The corresponding spectra, as obtained in the
simulations.}
    \label{F3}
\end{figure}
\newpara
We initially maintain the pump power in EDFA of $70$ mW and manipulate the
initial spectral phase by dint of the PS. The primary spectral
phase is flat, denoted as $[\alpha=0, L_{\mathrm{dis}}=0]$. In this
configuration, the soliton pulse passes through the PS without additional
spectral modulations. We conducted FROG measurements on the initial soliton
pulse. The reconstructed pulse duration $T_{0}$ is $\approx 810$ fs, with
the reconstructed spectral phase approximately following the second-order
distribution $\sim \omega ^{2}$, which corresponds to the second-order GVD.
This phase pattern results from the effects of the fibers and optical
elements in the setup, which can be compensated by adding an opposite second-order
spectral phase in the pulse shaper. Our tests indicate that the pulse duration is minimized
at the value of $\approx 352$ fs when the compensating GVD phase
corresponds to the dispersion length times a factor $\approx 1.7$ (\textbf{More details are produced in
Supplementary 4}).
\newpara
The next step involves fixing the spectral phase and recording the spectral
intensity (SI) by varying the pump power, i.e., effectively changing the
value of $B$ in Eq. \ref{SPM}. Figures \ref{F3}(a1) and (a2) display the recorded
SI for the initial pulse produced by PS with regular GVD. In
Fig. \ref{F3}(a1), the SPM-induced variations in SI are weak, due to the stretched pulse induced by the
offset GVD of the attached single-mode fiber. If the offset GVD is compensated, hence the pulse duration is
minimized, the SPM effect becomes evident, as shown in Fig. \ref{F3}(a2). Here, the interference dip
emerges around $290$ mW, corresponding to $B\approx 1.55\pi $ (\textbf{The relation between the
pump power and }$B$\textbf{\ is presented in supplementary 3}).
\newpara
As per the relationship between the number of lobes and the maximum phase
shift $(N-0.5)\pi $ \cite{agrawal2000nonlinear}, the theoretical value of $B$
in Fig. \ref{F3}(a2) should be $1.5\pi $. However, this threshold can be
lowered when the initial pulse is modulated by fractional GVD. For instance,
Figs. \ref{F3}(a3) and (a4) show the recorded SI for the loading phase with
parameters $[\alpha=1, L_{\mathrm{dis}}=2.42 L_{\alpha,\mathrm{FDL}}]$ and $%
[\alpha=0.2, L_{\mathrm{dis}}=0.88 L_{\alpha,\mathrm{FDL}}]$, respectively,
revealing a pronounced dip even at lower pump powers, $280$ mW ($B=$ $%
1.46\pi $) and $240$ mW ($B=$ $1.11\pi $), respectively. As $\alpha $
decreases, the deepest dip occurs at smaller values of $B$. Insets in these
panels show the recorded spectra for the pump power fixed at $240$ mW,
highlighting the enhancement in the pronounced dips induced by fractional
GVD. Moreover, the corresponding reconstructed temporal intensity(TI) for the
pump $240$ mW is shown at the bottom of Figs. \ref{F3}(a1)-(a4),
highlighting the profiles induced by fractional GVD, with the
``heavy tails".
\newpara
The observed spectrum displays a distinctive dip, which suggests the name of ``spectral valley" to this
feature. We designate the spectral valley by the symbol $\mathrm{\{N\}}$. In particular, $\mathrm{%
\{N=2\}}$ means that the spectrum has two peak lobes and one valley, as
shown in Figs. \ref{F3}(a3) and (a4). It is possible to produce the spectral
valley with multiple lobes ($\mathrm{N>2}$) by designing a phase structure
that incorporates multiple fractional segmented phases. Note that a Kelly
sideband in soliton pulses causes crosstalk in the spectrum, which was
eliminated by integrating a bandpass filter (see details in \textbf{%
Supplement 5}).
\newpara
To explain the observed spectral valley $\mathrm{\{N\}}$, we proposed a
``force" model by addressing the frequency chirp. Three
effective forces are referred to as ``attraction",
``repulsion", and ``equilibration". By
employing the force model, one can design the segmented phase to produce the
spectral valley of the $\mathrm{\{N\}}$ type. Details of the force model are
shown below in \textbf{Methods-2}.
\newpara
The required fractional spectral phase patterns, wrapped between $-\pi $ and
$+\pi $, are shown at the top of Fig. \ref{F3}(b1)-(b4). To generate the
spectral valley of \{2\}, we set FGVD to have LI $\alpha =1$:

\begin{equation}
 \alpha =1,\;L_{\mathrm{dis}}=3.87\times L_{\alpha ,\mathrm{FDL}}  \label{E7}
\end{equation}

Using the so-designed fractional phase, we have obtained the experimental SI
by varying the pump power, as shown in Fig. \ref{F3}(b1). At the pump power
of $\approx 250$ mW, two distinct peaks are observed in the spectrum. Figure %
\ref{F3}(c1) presents the spectrum versus $B$, as obtained from the
simulations of FNLSE (\textbf{see simulations in method-1}).
\newpara
To create the spectral valley \{3\} with three lobes, the segmented spectral phase
can be designed to carry a fractional and constant term, based on the force
model:

\begin{equation}\label{E8}
\left\{ \begin{array}{l}
\;\;\;\;\;\;\;\;\;\;\;\;\alpha=0, L_{\mathrm{dis}}=0\;\;\;\;\;\;\;\;\;\;\;\;\;\;\;|\lambda  - {\lambda _c}|
<  = 1.5 ~\textup{nm}\\ 
 {\alpha  = 1\;,{L_{\mathrm{dis}}} =  3.15\times {L_{\alpha, \mathrm{FDL}}}} \;\;\;\;\;\;\;
 \rm{otherwise}
\end{array} \right.
\end{equation}

where $\lambda _{c}$ is the central wavelength. Adding a global GVD phase to
this segmented phase pattern yields the final phase configuration, showcased
in the top plot of Fig. \ref{F3}(b2). When this phase is imposed on SLM,
three lobes are formed in SI with the increase of the pump power, as
displayed in Fig. \ref{F3}(b2), and by
simulations in Fig. \ref{F3}(c2), respectively. Notably, three lobes emerge
in both cases even for relatively small values of $B$. As the pump power
increases further, the two side lobes retain greater intensity than the
central one, subsequently fading out due to interference effects. This
behavior aligns with the force model, where the unique phase distribution
contributes to the emergence of three lobes: the lateral lobes are generated
by the repulsive force, while the central segment preserves the original
state due to the equilibration.

\newpara
Figures \ref{F3}(b3) to (c3) illustrate the scenario with the spectral
valley \{4\}, produced by dividing the phase configuration
into four segments, based on their effective-force interactions. This design
involves two fractional terms with LIs $\alpha =0.5$ and $1$, respectively.
Notably, the total force exerted by the segment with $\alpha =0.5$ must
be weaker than that produced by the segment with $\alpha =1$. Thus, the
required fractional GVD phase is expressed as follows:

\begin{equation}\label{E9}
\left\{
\begin{array}{ll}
 \alpha = 0.5, \; L_{\mathrm{dis}}= 0.79 \times L_{\alpha, \text{FDL}} & \quad |\lambda - \lambda_c|
 \leq 2 \; \text{nm} \\[8pt]
 \alpha = 1, \; L_{\mathrm{dis}} = 3.57 \times L_{\alpha, \text{FDL}}  & \quad \text{otherwise}
\end{array}
\right.
\end{equation}
\newpara
Upon applying the required spectral phase to the initial pulse, a four-lobe
pattern emerges within the pump power range of $70-250$ mW. Due to
interference effects, the two central lobes disappear but may reappear with
a further increase of the pump power. We do not show results for still
higher powers because of the plausible occurrence of the gain saturation in
that case\cite{ilday2004self}.
\newpara
More lobes can be realized by partitioning the fractional phase into more
segments. This entails designing a multi-functional force distribution,
encompassing attributes such as ``equilibration",
``weak repulsion", and ``strong repulsion".
To realize this, the spectral phase configuration is arranged as follows:
\newpara
\begin{equation}\label{E10}
  \left\{ \begin{array}{l}
\;\;\;\;\;\;\;\;\;\;\;\;\;\alpha=0\;, L_{\mathrm{dis}}=0, \;\;\;\;\;\;\;\;~~~|\lambda  - {\lambda _c}| <  = 1.6 ~\textup{nm}\\
{\alpha  = 0.5\;,{L_{\mathrm{dis}}} =0.97 \times {L_{\alpha ,\mathrm{FDL}}}}, \;\;1.6~ \textup{nm}<|\lambda  - {\lambda _c}| \leq 3.2~\textup{nm}\\
 {\alpha  = 1\;, ~{L_{\mathrm{dis}}} =3.57 \times {L_{\alpha ,\mathrm{FDL}}}}, \;\;\;\;\rm{others}. 
\end{array} \right.
\end{equation}

\newpara
The spectral phase defined in Eq. \eqref{E10} is displayed at the top of Fig. \ref{F3}(b4), where five segmented
phases are observed. Loading this phase into PS
and subsequently altering the pump power results in SI variations, as
displayed in Figs. \ref{F3}(b4). Experimental observations confirm that this
segmented phase design induces a spectral valley with five lobes.
\newpara
The simulations of SI (Figs. \ref{F3}(c1)-(c4)) maintain the
conservation of the input pulse energy, which is consistent with the experimental
settings. In the simulations and experiments alike, we
introduced an additional second-order-GVD phase term, with the size
corresponding to a factor $\approx 3.37$ times the dispersion length, as a
 global phase for the designed fractional-GVD pattern.  This global phase
serves a dual purpose: it counterbalances the added dispersion, caused by
the optical setup, and enhances the suitable temporal chirp for more
pronounced interference, resulting in a deeper dip. Moreover, at the pump
power level of $\approx 250$ mW, the exact number of the lobes in the
spectral valley can be found. This suggests a promising application for dense data encoding by using these spectral valley states, as shown in the next subsection.

\subsection{The applications of spectral valleys in  high-dimensional data encoding}
\begin{figure}
    \centering
    \includegraphics[width=18.5cm]{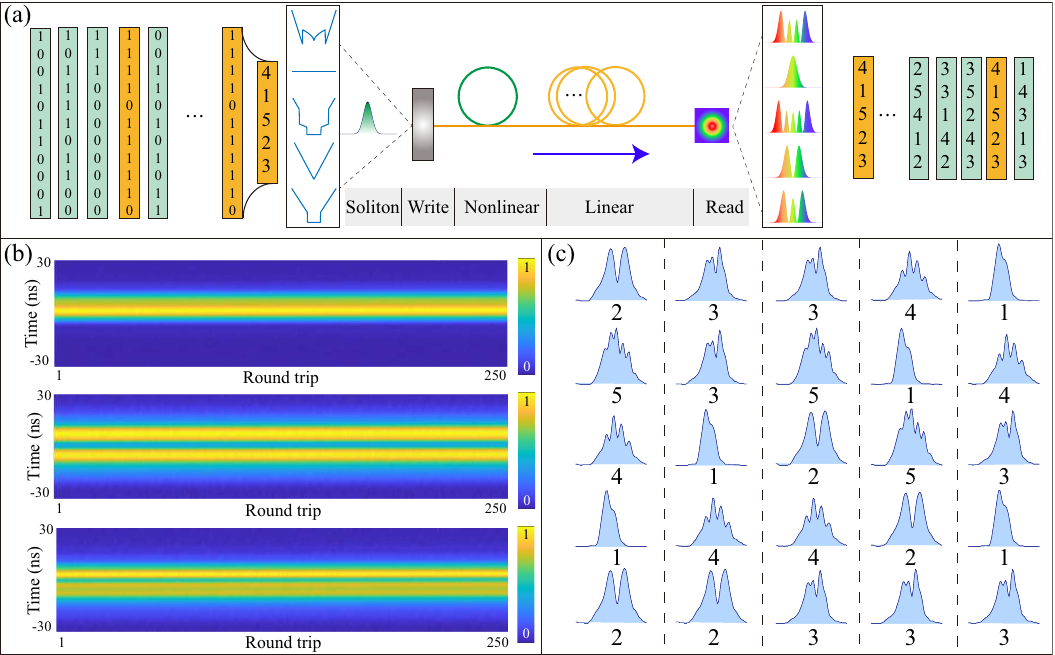}
   \caption{The application of the spectral valleys to high-dimensional data
encoding. (a): The pulse transmission regime for the data encoding from
binary to quinary by using five spectral-valley modes. It includes a `soliton'
pulse from a mode-locked fiber laser, the `write' section operated by a pulse
shaper, a `nonlinear' SPM section, a `linear' fiber link of the length $\simeq
100$ km, and a `read' section including a photodetector and oscilloscope. (b):
Stretched temporal profiles recorded by the oscilloscope for modes of $%
\mathrm{\{1\}}$, $\mathrm{\{2\}}$, and $\mathrm{\{3\}}$, respectively.
(c): The recorded temporal profiles of \{25412, 33142, 35243, 4152, 14313\}
strings, as realized with $25$ holograms and produced by the oscilloscope
for one round trip.}
    \label{F4}
\end{figure}

In experimental results presented in Fig. \ref{F3}, two relevant features
are observed. First, by adjusting the spectral phase suitably in the
hologram inserted into SLM, spectral valleys with multiple lobes can be
created. Second, these modes can be created with the same nonlinearity, or
the same pump power, which is beneficial and convenient for high-dimensional
encoding. Applying these spectral-valley modes to the optical signal
transmission is a promising possibility. Traditional optical transmission
systems usually use binary encoding, i.e., ``$0$" or
``$1$". However, creating five spectral-valley modes goes
beyond the binary format, allowing a quinary one. Figure \ref{F4}
demonstrates high-dimensional encoding with the format rearranging from
binary to quinary, by using five spectral valley modes $\{1\}$ to $\{5$%
\}, where $\{1\}$ stands for the fundamental mode without the spectral
split.
\newpara
To perform the data encoding, the first step is to transform the binary data
into a quinary format, shown in the left part of Fig. \ref{F4} (a). These
quinary datasets ($\{25412,33142,35243,41523,14313\}$) are then encoded in
sequence into the above-mentioned five spectral phases, which represent the
``write" process. In the second stage, the pump power is
fixed at around $250$ mW, and the corresponding spectral-valley mode is
generated under the action of SPM. The signal is then transmitted through $%
\approx 100$ km of the fiber (labeled ``linear" in Fig. \ref%
{F4}(a)) and registered by the photodetector and oscilloscope named the
``read" section. Due to the effect of the group delay
dispersion, induced by the long fiber, the stretched temporal profiles shown
by the oscilloscope are proportional to the spectral intensity \cite%
{mahjoubfar2017time,cui2021xpm,Liu2022SM}. At the last stage, the peak
numbers in these profiles are counted and the data is decoded back from
quinary to binary.
\newpara
Figure \ref{F4}(b) shows the stretched temporal profiles recorded on the
oscilloscope for the spectral valley modes $\{1\}$, $\{2\}$, and $\{3\}$,
respectively. The stretched temporal profiles demonstrate robustness over $%
250$ round trips and a dip is clearly observed at the center of the
profile. Figure \ref{F4}(c) shows one of the output profiles, with the
above-mentioned complex phase encoded by $25$ holograms. By counting the
number of peaks in these profiles, we obtain the same sequence. It should be
noted that the dips for the bifurcation modes of $\{3\}$ to $\{5\}$ are not
very deep, in terms of limitations imposed by the sampling rate and temporal
resolution of the oscilloscope. In addition, the higher-order GVD in the $100$-km
fiber affects the shape of these profiles. Nevertheless, the results
demonstrate that the spectral valleys offer promising potential applications
to optical signal processing, particularly for high-dimensional data
encoding.

\section{Conclusion}
In this work, we have theoretically explored and experimentally studied a
nonlinear LW (\textit{L\'{e}vy waveguide}) in the laser-cavity
configuration. The LW is modeled by the generalized FNLSE (fractional
nonlinear Schr\"{o}dinger equation), in which the pulse dynamics are
determined by the interplay between FGVD (fractional group-velocity
dispersion) and SPM\ (self-phase modulation). Our experiments conducted in
the intracavity regime reveals the existence of robust fractional solitons
featuring the characteristic ``heavy tails" at pulse edges.
\newpara
In addition, by implementing the extra-cavity scenario, we have built a
single-pass regime to emulate the spectral-temporal dynamics governed by the
generalized FNLSE. To explain the observed spectral valleys, we have
developed an effective ``force" model and performed its detailed analysis.
With the help of the force model, we have designed fractional segmented
phase patterns to produce multiple spectral valleys.
\newpara
 \SL{ The multiple lobes of the spectral valleys realized in the current study highlight two primary advancements. First, due to the beneficial interplay between the FGVD and SPM, these lobes can be observed even in the case of weaker nonlinearity. In contrast, for purely nonlinear systems, where regular SPM dominates, multiple lobes typically require a much higher degree of nonlinearity to be noticeable \cite{stolen1978self,agrawal2000nonlinear}. Second,  the lobes of the spectral valleys can be controlled on demand via the phase modulation, which offers a high-efficiency nonlinear shaping regime, in comparison with the linear pulse shaping methods by employing spectral amplitude modulations \cite{monmayrant2004new,weiner2011ultrafast}. }
These advancements indicate that the spectral valleys observed in this work are promising for applications to optical
data processing.
 Experimentally, we have demonstrated high-dimensional
optical data encoding using five-spectral-valley states, successfully
transmitting data over $100$ km of the single-mode fiber. Recent advances
also suggest the potential for developing a real-time signal processor with
a capacity of up to $17$ Tb/s, leveraging fractional-differentiation orders
\cite{tan2024photonic}, which warrants further investigation.
\newpara
\SL{Although the concept of the fractional derivatives and dispersion has a long history, their realization in physical systems, such as nonlinear fiber optics, is a relatively new and emerging topic. This inspires several promising perspectives: 1) The immediate goal is to explore additional solutions of the FNLSE. The current results focus primarily on positive fractional dispersion lengths, corresponding to the ``repulsion" case in the force model, which explains the observed spectral valleys. However, equally intriguing results are expected in the ``attraction" case, which would illustrate spectral squeezing. Recently two works have demonstrated an ``attraction" effect (narrowed spectrum) in similar contexts \cite{hoang2024observation,wang2024numerical}. By involving additional parameters, such as higher-order dispersion or stronger nonlinearity, even greater diversity in the pulse dynamics could be uncovered. 2) The second perspective is the realization of the spatial-temporal light synthesis by incorporating both fractional dispersion and fractional diffraction. This approach has significant potential for applications to optical encoding and spatiotemporal mode-locked laser architectures \cite{wright2017spatiotemporal,hui2023ultrafast}. 3) Finally, due to the mathematical similarity between the FNLSE in nonlinear optics and its quantum counterpart, this regime may serve as an effective model for emulating fractional quantum mechanics  \cite{laskin2000fractionalpath,laskin2000fractional}.
}


\section{Methods}
\subsection{Numerical Simulations for the FNLSE}

To simulate the intracavity regime as shown in Fig. \ref{F1}(d), we handle
three components separately: NLSE for the fiber, FGVD for the pulse shaper,
and the transfer function for SA (saturable absorber). For the fiber
segments and EDFA, the governing equation is

\begin{equation}
i\frac{\partial \Psi }{\partial z}=\left( \frac{\beta _{2}}{2}\frac{\partial
^{2}}{\partial t^{2}}+i\frac{g}{2}\right) \Psi -\gamma |\Psi |^{2}\Psi ,
\label{E-M1}
\end{equation}%
where we consider only the second-order dispersion, \SL{although high-order dispersion gives rise to more non-trivial results \cite{karpman1993radiation,vithya2018combined}}. The gain parameter $%
g=g_{0}\exp (-E(z)/E_{\mathrm{sat}})$ for the erbium-doped fiber is
determined by the small-signal gain $g_{0}$ and saturation energy $E_{%
\mathrm{sat}}$. The total energy of the pulse is $E(z)=\int_{-\infty
}^{+\infty }|\Psi(z,t)|^{2}\,dt$.

The pulse shaper, located in the free-space section of the cavity, is used
to engineer the additional fractional or regular dispersion,
which is modeled by the following map:

\begin{equation}
\tilde{\Psi}(\omega ,L_{\mathrm{4f}})=\tilde{\Psi}(\omega ,0)\sqrt{\delta }%
\exp \left(\frac{iDL_{\alpha ,\mathrm{dis}%
}|\omega |^{\alpha }}{2}\right) \mathrm{Rect}\left( \frac{\omega }{\omega
_{m}}\right) .  \label{E-M2}
\end{equation}%
Here, $L_{\alpha ,\mathrm{dis},}$ is the dispersion quantity for the given
GVD order $\alpha $ including the fractional one(s), which is (are)
characterized its (their) LI $\alpha $, and $L_{\mathrm{4f}}$ is the overall
length of the 4f pulse shaper. Parameter $\delta $ is the
overall transmission coefficient, and $\alpha =2$ represents the regular
(non-fractional) GVD. $\mathrm{Rect}(x)$ is the filter function of the pulse
shaper, where $\omega _{m}$ represents the maximum frequency range limited
by the size of the SLM.
\newpara
To achieve stable mode-locking, the setup also includes a SA
based on carbon nanotubes, as detailed in Ref. \cite{Liu2022SM}. The
absorber is characterized by its transfer function:

\begin{equation}
T_{\mathrm{ab}}=\sqrt{1-(\alpha _{\mathrm{ns}}+\alpha _{0}/(1+A^{2}/I_{%
\mathrm{sat}}\cdot A_{\mathrm{eff}}))}.  \label{E-M3}
\end{equation}%
Here, $\alpha _{\mathrm{ns}}$ is the unsaturated absorption, $\alpha _{0}$
is the linear limit of the saturable absorption, $I_{\mathrm{sat}}$ is the
saturation intensity, and $A_{\mathrm{eff}}$ is the effective cross-section
area of the fiber.
\newpara
Using Eqs. (\ref{E-M1})-(\ref{E-M3}), we simulated the evolution of the
pulse in the cavity. After approximately $100$ round trips, the output pulse
stabilizes, as verified by monitoring variations in its momentum and energy.
To confirm the long-term stability, the maximum number of round trips is set
to $1000$, and the final spectral and temporal amplitude profiles are
presented in Fig. \ref{F2}. The parameters used in these simulations are
summarized in Table \ref{tab:parameters}.

\begin{table}[htbp]
\centering
\begin{tabular}{|c|c|}
\hline
\textbf{Parameter} & \textbf{Value} \\ \hline
Central wavelength, $\lambda_c$ & 1560 nm \\ \hline
Maximum frequency range, $\omega_m$ & 54 ps$^{-1}$ \\ \hline
Length of SMF, $L_{\mathrm{SMF}}$ & 6.5 m \\ \hline
Dispersion of SMF, $\beta_{\mathrm{2,SMF}}$ & -21 ps$^2$/km \\ \hline
Nonlinear coefficient of SMF, $\gamma_{\mathrm{SMF}}$ & 0.0013 W$^{-1}$m$^{-1}$ \\ \hline
Length of Er fiber, $L_{\mathrm{Er}}$ & 2 m \\ \hline
Dispersion of Er fiber, $\beta_{\mathrm{2,Er}}$ & 12.1 ps$^2$/km \\ \hline
Nonlinear coefficient of Er, $\gamma_{\mathrm{Er}}$ & 0.003 W$^{-1}$m$^{-1}$ \\ \hline
Saturation energy, $E_{\mathrm{sat}}$ & 40 pJ \\ \hline
Dispersion parameter, $D$ & -21 ps$^{\alpha}$/km \\ \hline
Small-signal gain, $g_0$ & 5.5 m$^{-1}$ \\ \hline
Linear saturable absorption, $\alpha_0$ & 0.095 \\ \hline
Unsaturated absorption, $\alpha_{\mathrm{ns}}$ & 0.49 \\ \hline
Saturation intensity, $I_{\mathrm{sat}}$ & 21 MW/cm$^2$ \\ \hline
Transmission rate, $\delta$ & 0.2 \\ \hline
Output coupling ratio, $R_{\mathrm{out}}$ & 0.10 \\ \hline
Effective cross-sectional area, $A_{\mathrm{eff}}$ & $63.6 \times 10^{-12}$ m$^2$ \\ \hline
\end{tabular}
\caption{Simulation parameters for the intra-cavity regime}
\label{tab:parameters}
\end{table}

The simulations of the extra-cavity regime are similar, in which case Eq. (%
\ref{E-M2}) is first used to imprint the initial spectral phase onto the
incident pulse. Then, Eq. (\ref{E-M1}) governs the propagation.

\subsection{The force model to explain the FGVD-SPM interplay}
\begin{figure}[htbp]
    \centering
    \includegraphics[width=18.5cm]{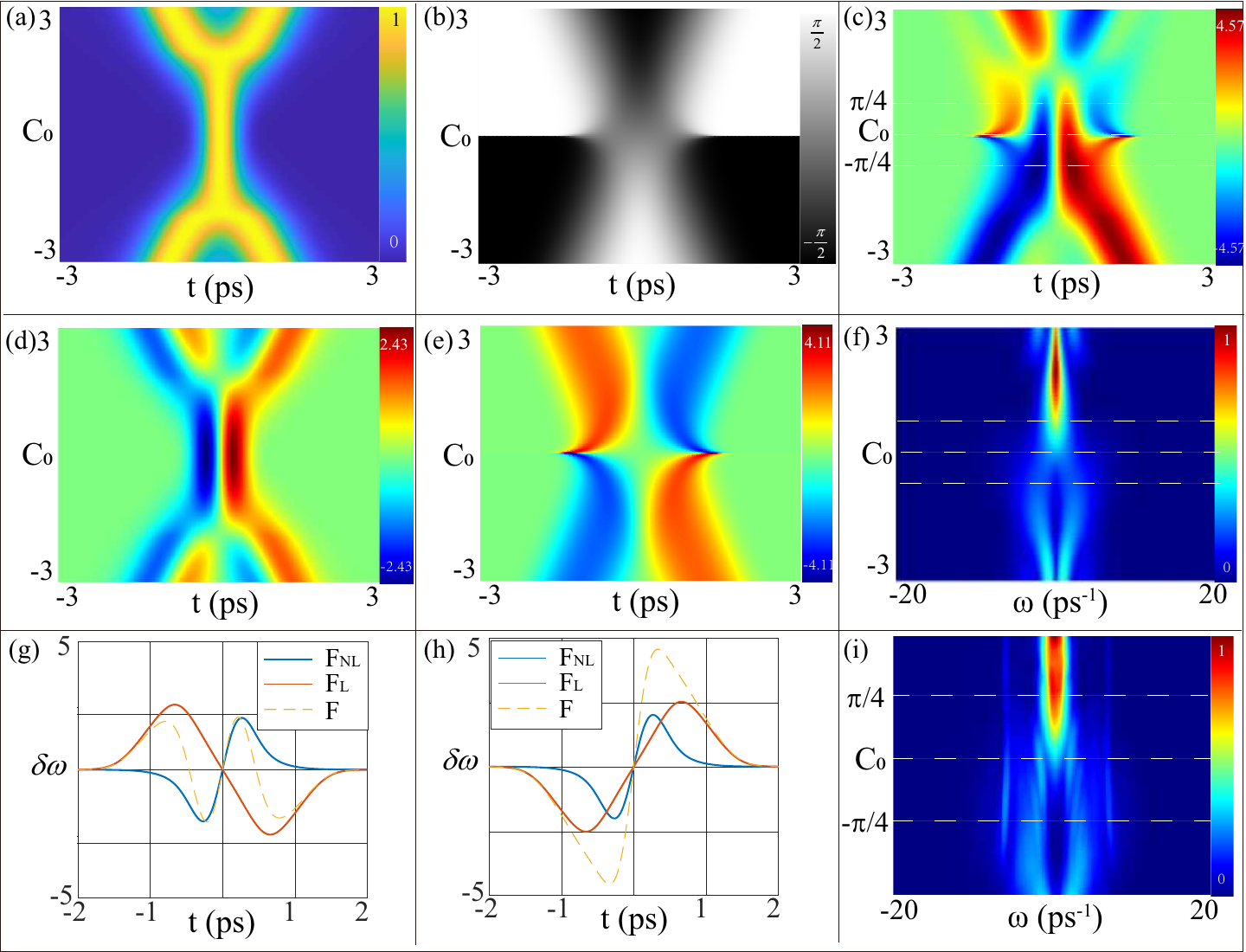}
   \caption{Simulations and experiments for pulse's spectra under the action of
FGVD and nonlinearity. (a) and (d): The normalized temporal intensity and {gradient ($%
-\partial I/\partial t$). (b) and (e): The corresponding temporal phase and
gradient ($-\partial \protect\phi /\partial t$)}. (c) The summary effective
force for $B=\protect\pi /2$. (f) The calculated spectral intensity for $C_{0}$ varying
between $-3$ to $+3$. (g)-(h): The calculated three forces for $C_{0}=+\protect\pi %
/4$ and $C_{0}=-\protect\pi /4$, respectively. {(i) The measured spectral intensity for $%
C_{0}$ varying from $-1.8$ to $1.8$.}}
    \label{F5}
\end{figure}

The dynamics of the spectral valleys are similar to the oscillating
structure induced by standard SPM. However, the unique feature observed in
the present setting is the \textit{direct} transitions from $\{1\}$ to
\{N\}, while, as mentioned above, solely \textit{cascaded} transition
sequences occur with the increase of the nonlinearity in the case of
standard SPM. To study the FGVD-SPM interplay, we focus on the frequency
chirp,
\begin{equation}
\delta \omega (B,t)=-\frac{\partial \lbrack \phi _{NL}(t)+\phi _{L}(t)]}{%
\partial t}=F_{NL}(t)+F_{L}(t),  \label{chirp}
\end{equation}%
where $\phi _{L}(t)$ denotes the temporal phase induced by FGVD, and $\phi
_{NL}(t)$ is the SPM-induced nonlinear phase. The derivatives $%
F_{NL}(t)=-\partial \phi _{NL}(t)/{\partial t}\equiv -B\partial I(t)/{%
\partial t}$ [see Eq. (\ref{SPM})] and $F_{L}(t)=-\partial \phi _{L}(t)/{%
\partial t}$, which may be compared to the effective forces that induce the
frequency shift in the input pulse \cite{zheltikov2023lagrangian}, are
represented by the total effective force, $F(t)$. This picture gives rise to
three distinct scenarios, which are defined as ``repulsion" (%
$F(t)>0$), ``attraction" ($F(t)<0$), and ``equilibration", ($F(t)=0$). In particular, the ``repulsive"\
force leads to an increase in the frequency, akin to the blue shift of optical
pulses.
\newpara
To visualize the ``force" under the action of FGVD with $\alpha=1$, we consider the initial pulse with the Gaussian spectral
shape, \textit{viz}.,
\begin{equation}
\tilde{\Psi}(\omega )=\exp \left[ -\left( \frac{\omega }{\Delta \omega }%
\right) ^{2}-iC_{0}|\omega |\right] .  \label{ES5}
\end{equation}%
Here, $\Delta \omega $ is the bandwidth and $C_{0}$ represents the strength
of the fractional GVD. The respective temporal field can be written as
\begin{equation}
    \Psi (t)={\mathcal{F}^{-1}}\left[ {\tilde{\Psi}(\omega )}\right]=\frac{\Delta \omega \sqrt{\pi}}{2}\left({{{E_{n}}\cdot (1-i{f_{n}})+{E_{p}}\cdot (1-i{f_{p}})}}\right),
\end{equation}\label{ES6}
with
\begin{equation}
\left\{
\begin{array}{l}
{f_{n}}=\mathrm{erfi}\left( \frac{{\Delta \omega (C_{0}-t)}}{2}\right), \\
{f_{p}}=\mathrm{erfi}\left( \frac{{\Delta \omega (C_{0}+t)}}{2}\right), \\
{E_{n}}=\exp \left( {-\frac{{\Delta {\omega ^{2}}{{(C_{0}-t)}^{2}}}}{4}}
\right), \\
{E_{p}}=\exp \left( {-\frac{{\Delta {\omega ^{2}}{{(C_{0}+t)}^{2}}}}{4}}%
\right),
\end{array}%
\right.
\end{equation}
where $\mathrm{erfi}(x)=\int_{0}^{x}\exp (\tau ^{2})d\tau $.
Based on \eqref{ES6}, one can
calculate the temporal phase induced by FGVD and the SPM-induced nonlinear
phase, which makes it possible to obtain the force distributions, i.e., $%
F(C_{0},t)$. The calculated normalized temporal intensity and phase are
shown in Figs. \ref{F5}(a) and (b), and the corresponding gradients are
shown in Figs. \ref{F5}(d) and (e), in which we set $\Delta \omega =4$ $%
\mathrm{ps^{-1}}$ and $B=\pi /2$.

\newpara
By taking these findings into account, we obtain the total force $F(C_{0},t)$%
, which is shown in Fig. \ref{F5}(c). It is clearly seen that two branches
appear with a strong dip at $C_{0}<0$, which implies that the repulsive
force creates a spectral valley. This conclusion is corroborated by the
spectrum, as shown in Fig. \ref{F5}(f) (the simulation) and Fig. \ref{F5}(i)
(the experiment). Specifically, Figs. \ref{F5}(g) and (h) show the force
distribution for $C_{0}=+\pi /4$ and $-\pi /4$, respectively. In the latter
case, the two forces have identical signs, enhancing the total
``repulsion" force. On the other hand, the two forces almost
cancel each other for $C_{0}=+\pi /4$, showing the ``equilibration".

\medskip
\newpara
\textbf{Supporting Information} \par 
Supporting Information is available from the Wiley Online Library or from the author.

\medskip
\newpara
\textbf{Acknowledgements} \par 

The authors would like to thank Prof. Ebrahim Karimi's SQO group for providing the spatial light modulator. We appreciate the following colleagues in Polytechnique Montr\'{e}al for their help in experiments and discussions:  Godbout Nicholas, Mikael Leduc, Sho Onoe, Laurent Rivard, Gabriel Demontigny, Patrick Cusson, Marco Scaglia, \'{E}mile Dessureault, Gr\'{e}gory-Samuel Zagbayou, Rodrigo Itzamna Becerra Deana, as well as \'{E}mile Jetzer.  In particular, we thank Prof. Kai Wang's (McGill University) beneficial discussions on the results and the potential regimes, Chi Zhang's (Polytechnique Montr\'{e}al) help in the design of the saturable absorber, as well as Prof. Yudong Cui's (Zhejiang University) discussions concerning mode-locked fiber lasers. In addition, we appreciate the assistance of Christine Tao (Tempo Optics Inc.) in the improvement of the figure quality.

\medskip
\newpara
\textbf{Funding}

Natural Sciences and Engineering Research Council of Canada (NSERC) via the Canada Research Chair (CRC) of D.V.S.; the Fonds de Recherche du Qu\'{e}bec Nature et Technologies (FRQNT) via Institut Transdisciplinaire d' Information Quantique Transdisciplinary Institute for Quantum Information (INTRIQ); PBEEE/Bourses de court s\'{e}jour de recherche ou perfectionnement from FRONT; Canada Research Chair Program (CRC) of E. K., NRC-uOttawa Joint Center for Extreme Quantum Photonics (JCEP); Mitacs Accelerate Program; Israel Science Foundation via grant No. 1695/22 by B. A. M. European Union's Horizon Europe research and innovation programme under grant agreement No. 101070700 (project MIRAQLS).
\bibliographystyle{MSP}

\end{document}